\documentclass[manuscript]{acmart}
\usepackage{tcolorbox}
\usepackage{tabularx}

\begin{document}
\AtBeginDocument{%
  \providecommand\BibTeX{{%
    Bib\TeX}}}

\setcopyright{acmlicensed}
\copyrightyear{2026}
\acmYear{2026}
\acmDOI{XXXXXXX.XXXXXXX}
\acmConference[Under Submission]{}{April 2026}{New York, NY}
\acmISBN{978-1-4503-XXXX-X/2018/06}




\title{fog: Generating Expressive Motion through Layered Function Composition of Code for Heider-Simmel Animation}

\author{Vivian Liu}
\email{vivian@cs.columbia.edu}
\affiliation{
  \institution{Columbia University}
  \city{New York}
  \state{NY}
  \country{USA}
}

\author{Lydia Chilton}
\email{chilton@cs.columbia.edu}
\affiliation{
  \institution{Columbia University}
    \city{New York}
  \state{NY}
  \country{USA}
}


\renewcommand{\shortauthors}{Anon et al.}
\newcommand{\vivian}[1]{{\color{purple} author: #1}}

\begin{abstract}
Motion and emotion are core parts of intelligent, expressive behavior. In this paper, we introduce \textit{fog}, a layered function composition framework for generating motion functions. We demonstrate how \textit{fog} can be used to express motion and emotion in Heider-Simmel style animations. This code generation framework can help users generate functions for verbs, adverbs, and emotions to capture the trajectory, manner, and feeling of motion. It is complemented by an animation editor that helps users provide spatial context in the form of marks, arrows, and paths. We evaluate our approach with a technical evaluations investigating different function generation and composition setups. In a mixed-methods user study (n=12), we find that \textit{fog} can support users can support users by relaying spatial context they mark and draw and giving them more dimensions for control.

\end{abstract}

\ccsdesc[300]{Human-centered computing~Natural language interfaces}
\ccsdesc[500]{Applied computing~Media arts}
\ccsdesc[100]{Applied computing~Psychology}
\ccsdesc[300]{Human-centered computing~User interface programming}

\keywords{motion, emotion, animation, function composition, generative AI, code generation, human-AI interaction, Heider-Simmel, generative user interfaces}
\begin{teaserfigure}
  \includegraphics[width=\textwidth]{figures/fog_teaser_v2.pdf}
  \caption{In \textit{fog}, users utilize LLMs to generate motion and emotion functions by implementing class definitions for verbs, adverbs, and emotions. These functions can compose with each other and act over an entity's internal and external state (energy, velocity, collision, paths) to show behavior. The Heider-Simmel animation recreated in \textit{fog} shows social motion (chase), motion decoration (aggression), and emotionally-expressive motion (fear), and social dynamics (rejection and protection).}
  \Description{The teaser figure shows the four states of the Heider-Simmel animation recreated using fog's framework. There are four stills representing the last frame of each state. Each state shows the entities: a big aggressor triangle (red), a yellow circle, and a smaller blue triangle. The motion trails on these entities is also shown. Below the canvas stills each entity is shown with its relevant primary and secondary motion layers. Calm (state 1) reads out: yellow circle drifts, blue triangle follows yellow circle, red triangle enters. Chase (state 2) reads out: aggressively chases, fearfully avoids. With fearfully and aggressively as the only instance of secondary motion. The third state Intervene shows that the yellow circle guards the blue triangle at a specific mark the user makes on the canvas. The fourth state shows the red triangle exiting on a path. We also color code the layers for if they are primary, secondary, or laden with user drawn spatial context (marks, paths).}
  \label{fig:teaser}
\end{teaserfigure}


\maketitle

\section{Introduction}

Motion is emotion without the ‘e’. As people, we tend to interpret meaning, intentionality, and animacy from things that move. This has been well established in cognitive science. In 1944, Heider and Simmel \cite{HeiderSimmel} ran a famous psychology experiment using an animation where three shapes played out a series of events. They found that even when the animation ran without sound or words, participants still described a story of love, joy, fight, and chase. 

Heider and Simmel created their animation scene in 1944 using stop-motion, but today users can create these animation scenes by generating animation code. Generative AI tools have proven to be powerful at animation because within code representations, AI can manipulate the storyline of the animation as well as how it is visually expressed in terms of physics and paths. However, control with generative AI for animation editing is still very stochastic. Users struggle with to articulate the animation in their mind and condense it down into a prompt. Furthermore, their prompts can only get so granular in describing speed, timing, ordering, and paths. AI tools need representations that unlock spatial and deterministic controls for users. Users want to be able to point to where a character should go, to draw the paths they take through the scene, and to isolate on the manner and feeling of the motion.


The problem statement we address in this paper is how we can help novices author motion with multi-faceted control over its narrative, spatiotemporal, and numerical aspects. To make this problem statement concrete, we use Heider-Simmel type animations as our research context. These shape stories are simple enough to computationally represent, yet can represent complex social dynamics due to the human tendency to anthropomorphize.  They allow us to study what are the primitives and building blocks necessary for control during animation editing. To theoretically ground our approach, we draw upon theories from graphics (Reynold's model of autonomous agent behavior) \cite{Reynolds:2002:Steering}, cognitive science (semantic space theory) \cite{Tversky:05:VisuospatialReasoning}, and animation (Disney principles) \cite{Disney:81:PrinciplesAnimation}. Our main insight is that we can create computational representations of motion and emotion and implement them through code generation. Both motion and emotion can be represented by forces, energy, and states in time -- things AI can now tractably generate due to advancements in logical and mathematical reasoning.  We can create a hierarchy of code representations that at the lower-level manage the numerical aspects of external (motion) and internal (emotion) state, and at the higher-level combine in layers.




We present \textit{fog}, a function-composition framework that creates a hierarchy of \verb|Motion| primitives that combine in primary and secondary layers. These primitives span \verb|Verbs|, \verb|Adverbs|, and \verb|Emotions|. For example, entities can \verb|chase| and \verb|avoid| each other, instancing \verb|Verbs| to create attraction and rejection in motion. \verb|Adverbs| and \verb|Emotions| can compose with these \verb|Verbs| to modify motion performance. In Fig.~\ref{fig:teaser}, we show how a larger red triangle \verb|aggressively(chases)| a smaller blue triangle, who \verb|avoids fearfully|. In the next state, the yellow circle intervenes to \verb|guard| the blue triangle, who \verb|clings| to it. User compose functions by combining the \verb|Verb|, which extends \verb|PrimaryMotion|, with \verb|Emotions| and \verb|Adverbs|, which extend \verb|SecondaryMotion|. The former defines the main trajectory and action, while the latter decorates with another layer of motion dynamics.




We evaluate \textit{fog} in two parts. The first evaluation looks at different function generation and composition setups to see how imposing structure affects quality and composability of generated motion. The second evaluation, a user study, looks at \textit{fog}'s ability to support editing and interaction. We conclude with a discussion of how the abstract class paradigm introduced by \textit{fog} can be a new surface area for prompt-based interaction and how interfaces can balance prompts and GUI controls.

Our contributions are the following: 
\begin{itemize}
\item \textit{fog}, a function composition framework to express \verb|Motion| and \verb|Emotion| through code generation. It introduces a hierarchy of \verb|Primary| and \verb|Secondary Motion| classes that span \verb|Verbs|, \verb|Adverbs|, and \verb|Emotions|.

\item \textit{fog} as an animation system, which shows how users can 1) generate new Primary and SecondaryMotion functions on-the-fly and 2) use both prompts and direct manipulation for editing.

\item Technical evaluation showing that imposing more structure during code generation produces a tradeoff with expressive range.    

\item Mixed-methods user study (n=12) showing that \textit{fog} opens up new affordances for spatial control and was preferred over a baseline for its overall support in control.

\end{itemize}

\section{Related Work}

\subsection{Theories of Motion and Early Work in Animation} 

Our ability to interpret animacy and intent from motion is the core result of the Heider-Simmel paper \cite{HeiderSimmel}. In this experiment, participants described how in an animation of only simple shapes, they could see a story play out where a bigger triangle bullies two smaller shapes. Heider-Simmel interpreted these results as evidence that people can understand causal origins of motion, even when characters are highly abstract. Cognitive psychology has also shown that from a young age, we know the difference between animate and inanimate motion \cite{Gelman07AnimateInanimate, Tversky:05:VisuospatialReasoning}. Animate motion is motion attributed to living beings, while inanimate motion is motion driven by physics acting on objects (e.g. billiard ball collisions). Getting motion right is a balance between getting the physics right and making the motion look intentional. Disney animators proposed principles like composing primary and secondary motion, squash-and-stretch, and arcs to provide technical guidance on how to create this ``illusion of life" \cite{Disney:81:PrinciplesAnimation}.

Early work in autonomous and intelligent motion happened in the 1980's. Reynolds in his influential paper on steering behaviors \cite{Reynolds:2002:Steering} proposed that human behavior can be considered a “hierarchy of motion behaviors”. By creating a velocity- and force-based model, he showed that entities could be animated to flee and seek towards targets and to flock and swarm. Other works like Improv \cite{improv} and A Behavioral Language \cite{ICT:ABL} also introduced frameworks for hierarchical motion planning with different degrees of freedom. However, these frameworks tended to focus on physics-informed motion and not on the emotionally-expressive aspects of motion. Of note is also Herb Simon's framework \cite{HerbSimon:1967:Motivational} on the motivational and emotional controls of cognition; while he does not touch upon motion in specific, he describes how intelligent, goal-seeking behavior cannot be disentangled from drives like feeling, or ``hot cognition".

\subsection{Computational Theories of Emotion}

There are established models for the computational representation of emotion. The first is Russell's circumplex model, which helped popularized the two-axis understanding of emotion as valence and arousal \cite{Russell:1980:Circumplex}. This model, as well as Mehrabian's Pleasure-Arousal-Dominance model \cite{Mehrabian:1996:PAD}, state that emotions are often combinations of ``pleasure and arousal components". More recently, Cowen proposed the Semantic Space Theory of Emotion \cite{Cowen:SemanticSpaceTheory}. This theory states that emotions are “high-dimensional" states that can be computationally represented in a continuous vector space. In such a space, emotions can blend; an emotion like ``bittersweet” can arise as a combination of happiness, sadness, and nostalgia. 

Other early works in computer animation did focus on emotive and expressive performance \cite{Kahn:79, Badler:85}. Strassman’s ``Desktop Theater” system \cite{Strassmann:1991:DesktopTheater} converted natural language into animation by storing a library of motion behaviors and keeping both numerical and symbolic representations of an animation. This allowed the system to animate characters to express \textit{anger} or move \textit{rudely}. A key idea in this work and others at the time \cite{Unuma:95:Fourier} was that emotion was a mood or personality-dependent factor \cite{Chi:2000:EMOTE, Durupinar:2016:PERFORM}, a resource that would affect how the motion was performed. 


\subsection{AI Systems for Motion}

More recent work has also looked at semantically meaningful and emotionally expressive motion. Geppetto \cite{Desai:2019:Geppetto} is a pre-LLM machine learning system that helps users design expressive robot motion, like whether it moves happily or sadly. Since the LLM-era, many systems \cite{Angert:2023:Spellburst, Tseng:2024:Keyframer, Logomotion} have demonstrated that AI can enable novices to animate through prompts and code generation. However, systems have also established that canvas interactions are important for animation. Narrative Motion Blocks \cite{Bourgault:2025:Narrative} demonstrates how direct manipulation can be bound to prompts. The work closest to ours is Toyteller \cite{Chung:25:Toyteller}, which also looks at Heider-Simmel animations. They train a transformer and learn a representational space between language and motion but do not look at animation from a code generation perspective.

\textit{fog} synthesizes insights from these bodies of work to contribute a framework for code generation that semantically represents motion and emotion. 

\section{Formative Work}

Animating emotions with only simple shapes in the style of Heider-Simmel animations is not the most common animation problem. To understand how to approach it better, we conducted a few formative steps. First, we talked to professional animators to understand what their approach would be, what tools they use today, and how their workflows are evolving with AI.


\subsection{Formative Interviews with Professional Animators }

\subsubsection{Methodology} We conducted three semi-structured interviews with animators. We reached out to animators on Upwork through purposive sampling. We interviewed three professionals (E1-E3) who were chosen based on their extensive portfolios of motion graphics and animation work. E1 had experience in After Effects, Cinema4D, as well as AI design tools. E2 was a designer who primarily worked on logos and did animations upon client request. They had experience with After Effects and some exposure to other animation tools like Rive, Figma, and Framer for animation. E3 was a character animator who worked primarily in Toon Boom Harmony. 
The interviews were semi-structured: first, we had professionals walk through one or two projects in their portfolio to talk through their workflow and tools; then, we asked them how they would approach the animation of a simple shape to express emotion. In total, interviews lasted approximately 1-1.5 hrs, and professionals were compensated \$40 for their time and expertise.

\subsubsection{Findings}
First, animators were experimenting with AI tools through custom scripts. E1 and E3 all had experimented with tools like ChatGPT and Claude to create custom programs that would help them with friction points in their workflow. For example, E3 had asked Claude to create a script that could fill in inbetween frames for them. E1 took scripts from ChatGPT and pasted them in as custom timeline expressions in After Effects to show how they could get the system to support rigidbody physics where it could not before.

Second, animators often searched for professional presets. Even when animators are professionals, they often have to reach for presets. E2 described how most of their recent work for clients was focused on creating a polished and premium motion feel. To get that, they often used components shared in communities. While these components worked great for E2 in isolation, they could not be easily combined – they lacked the shared state and tool awareness to support composition.

Third, during the exploratory task, we found that animators made use of everything from paths, speed, energy, and deformation to animate emotion on a simple shape. When asked to animate sadness, E3 sketched out a shape and explained how they would use exaggeration and a droopy deform to make a circle look sad, \textit{“It needs a sort of exaggeration in the opposite direction of the movement before it starts moving or else it looks robotic. So I always do a slight exaggeration since it's going to droop downwards and flatten.”} They would keyframe and animate using a combination of squash-and-stretch, exaggeration, and anticipation. They also mentioned that while they expressed sadness in one way (with deformation), they could express another emotion like anger with completely different channels. \textit{“Anger is a stiffer emotion. It can rattle around and be flying around the screen to look angry.”}

These interviews motivated our final set of goals for \textit{fog}: 
\begin{itemize}
\item [DG1] \textbf{Provide people with a basic motion vocabulary, but make it an open set so users can dynamically generate new primitives.} The \textit{fog} framework should intuitively match the vocabulary people have for motion, such as verbs, adverbs, and emotions. 

\item [DG2] \textbf{Enable function composition.} One of the fundamental principles of animation is the split between primary motion and secondary motion. \textit{fog} should help users create layers of primary and secondary motion that can combine to express plot and emotion. Let verbs, the primitive of action in natural language, own \textit{trajectory} of the motion. Let modifiers like adverbs and emotions own the \textit{manner} and \textit{feeling} of the motion.

\item [DG3] \textbf{Give users a ladder of controls that span semantic (higher-level, storytelling controls) to perceptual refinement (lower-level, granular controls).} Animators noted that tweaking exaggeration and speed are key for motion refinement. At the highest level of this ladder, give users semantic and prompt controls, which can make more sweeping changes. In the middle level of this ladder, functions and motion tools can be intermediate layers, so users can have structured, local exploration. In the lowest level, give users perceptual controls so they can manipulate hard-to-describe things like speed and timing.

\item [DG4] \textbf{Give users methods for passing in spatial context to AI.} All animators mentioned that positioning, blocking, and trajectory were imperative to get right and iterate upon. \textit{fog} should provide users with marks, arrows, and paths. These should be integrated as spatial context with prompt-based interaction, so people have ways to point, draw, and describe spatial relationships to AI.

\end{itemize}

\section{System}

\subsection{The \textit{fog} Framework}
\begin{figure}[h]
        \centering
        \begin{tcolorbox}[
        colframe=gray, 
        colback=white,     
        arc=4pt,           
        boxrule=1pt,       
        halign=center,
        left=2pt, right=2pt, top=2pt, bottom=2pt 
    ]\includegraphics[width=\textwidth]{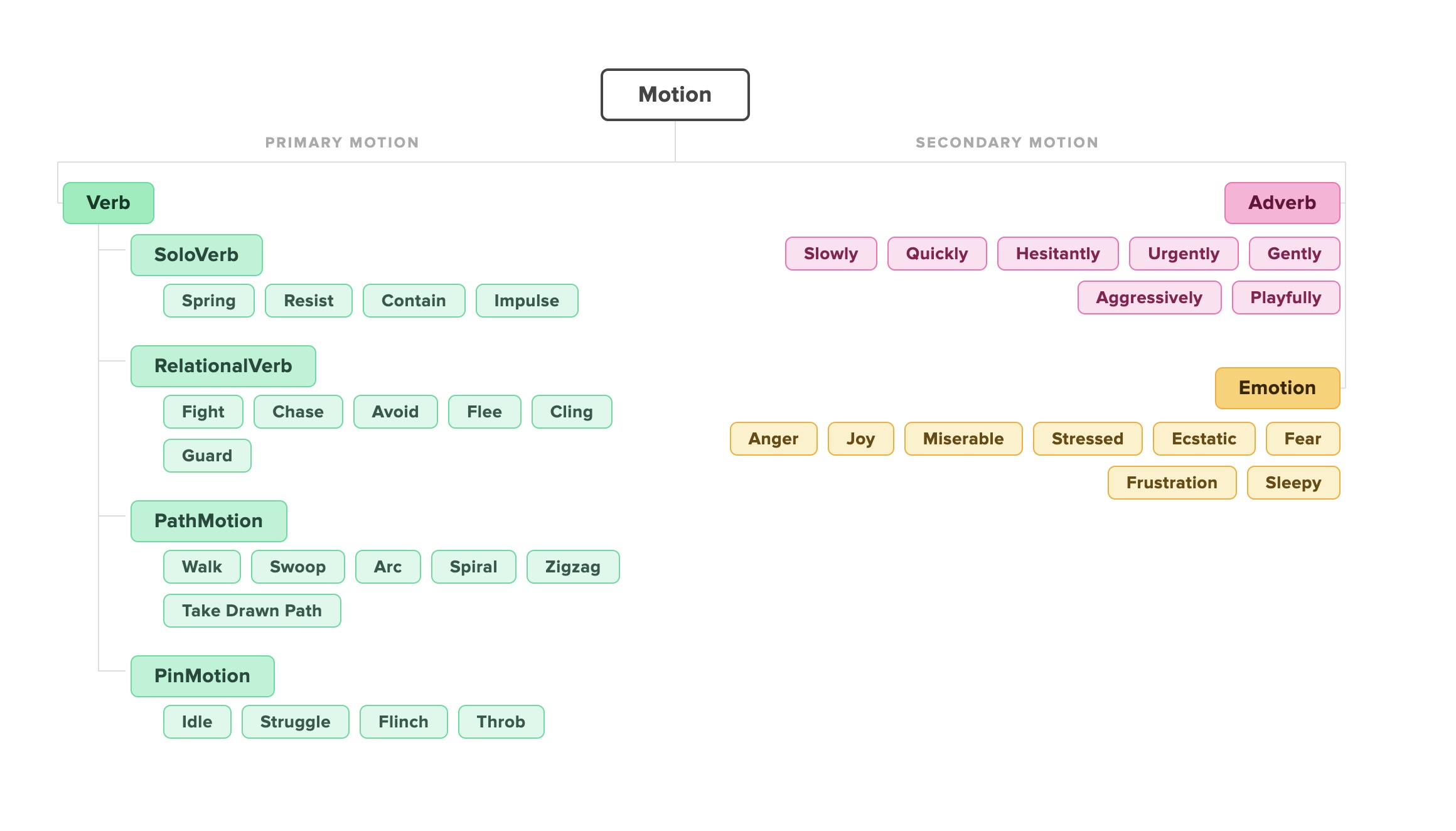} 
    \end{tcolorbox}
        \caption{\textit{fog}'s motion layer class hierarchy, which creates layers of abstraction by connecting words in language (verbs, adverbs, and emotions) to generated motion. Motion is an abstract class that splits into PrimaryMotion (Verbs) and SecondaryMotion (Adverbs, Emotions). Leaf nodes are examples of the class / concrete implementations.}
        \Description{ A stylized tree diagram shows that Motion splits out into Primary and Secondary Motion on the left and right. Verb headers the left side, splitting out into SoloVerb, RelationalVerb, PathMotion, and PinMotion. Beneath these subheaders, there are examples of different verbs. On the right we have Adverb, beneath which there are leaf nodes like Slowly, Quickly, Aggressively. We also have Emotion, which splits out into leaf nodes like Anger and Joy.}
        \label{fig:fog_class_hierarchy}
    \end{figure}

\textit{fog} is a hierarchy of abstract classes for expressing motion and emotion. These abstract classes include \verb|Scene|, \verb|State|, \verb|Entity|, \verb|Emotion|, \verb|Motion|, \verb|PrimaryMotion|, \verb|SecondaryMotion|, \verb|Verb|, \verb|Adverb|, and \verb|Emotion|. These classes bridge the higher-level symbolic aspects of motion with its lower-level numerical underpinnings (components like velocity, acceleration, intensity, and exaggeration). The class hierarchy is illustrated in Fig.~\ref{fig:fog_class_hierarchy}.

\begin{figure}
\centering
\begin{tcolorbox}[
        colframe=gray, 
        colback=white,     
        arc=4pt,           
        boxrule=1pt,       
        halign=center
    ]          
\includegraphics[width=0.75\textwidth]{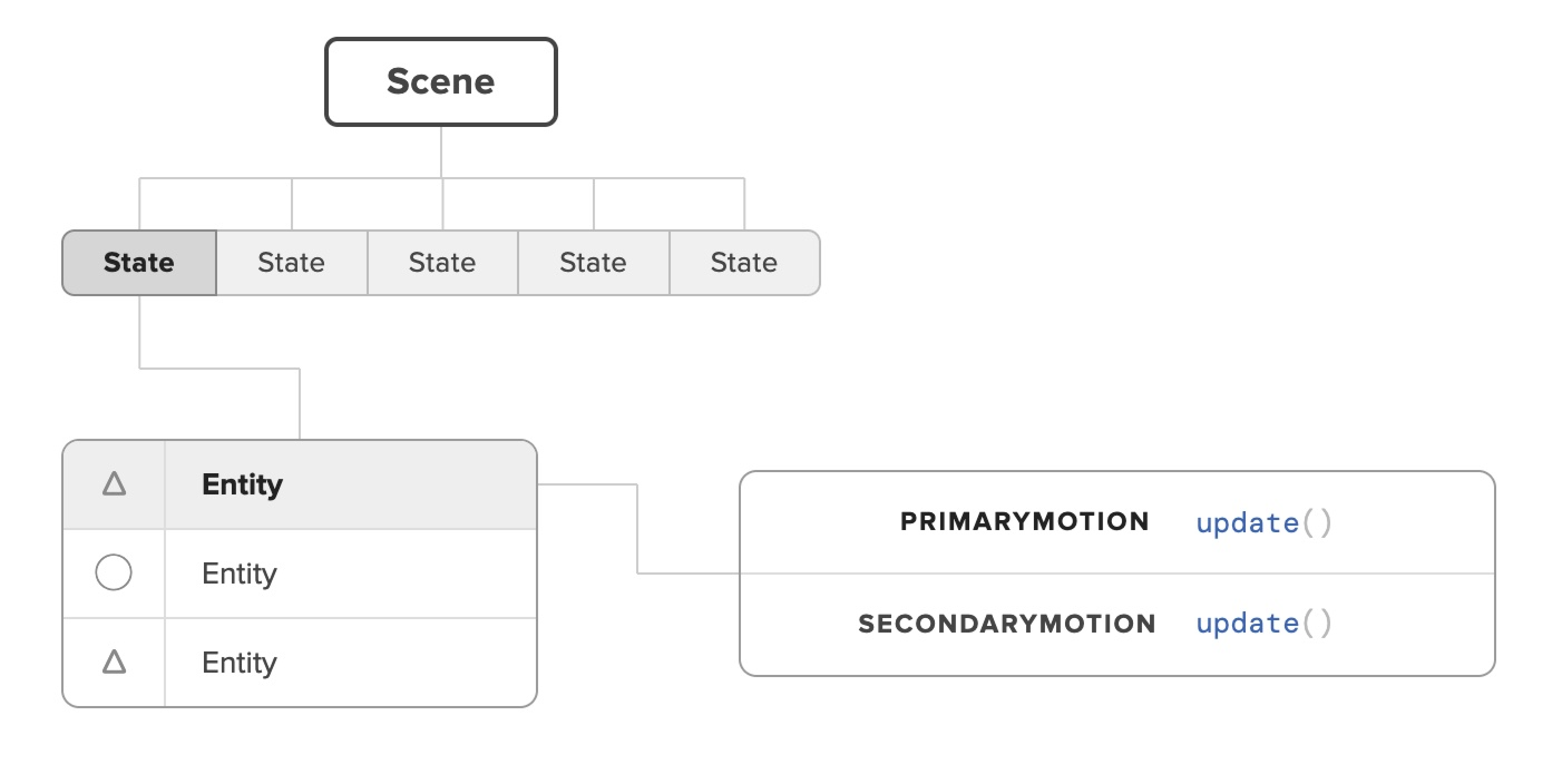}
\end{tcolorbox}
        \caption{\textit{fog}'s representation of an animation scene, broken down in terms of States, Entities, and Primary Secondary motion layers that compose.}
        \Description{Scene splits out into a timeline of states. Each state has a group of entities. Each entity has a primary motion and a secondary motion layer. Primary fulfills update(), and secondary motion layers fulfill update(). We depict this schematically. }
        \label{fig:fog_scenestate}
    \end{figure}

\subsubsection{Scenes, States, Entities} \verb|Scenes| are the highest-level representation of the animation. They consist of \verb|States|, and each \verb|State|  describes the animation on \verb|Entities|. The \verb|Entity| class represents characters as shapes (circles, triangles) and keeps track of properties like position, size, velocity, energy, position history, and orientation. To give entities a sense of animacy, we apply squash-and-stretch in the direction of their velocity and give them motion trails (a tail). Each entity accumulates acceleration into velocity and integrates velocity into position. For example: \verb|entity.x += entity.vx * dt|.


\subsubsection{Function Composition Execution Model}
In each state, entities can have one \verb|PrimaryMotion| function and multiple \verb|SecondaryMotion| functions. This split is motivated by Disney's principle of primary and secondary motion \cite{Disney:81:PrinciplesAnimation}. \verb|PrimaryMotion| layers implement \verb|update()|, which describes how the entity should update per-frame in the state. \verb|PrimaryMotion| describes the primary trajectory and meaning of the motion, like something a verb or path would describe. \verb|SecondaryMotion| describes how secondary action, manner, and feeling can modify the \verb|PrimaryMotion| and layer on top of it.

SecondaryMotion and PrimaryMotion compose with each other as layers. This makes it possible for an end user to swap between different layers of motion while holding one quality (trajectory for \verb|PrimaryMotion|, manner / feeling for \verb|SecondaryMotion|) of the entity motion constant. This layering of functions is the function composition ($f(g(x))$) that gives our framework its name -- \textit{fog}. The state machine and execution model are illustrated in Fig.~\ref{fig:fog_scenestate}. 





\subsubsection{Verbs as PrimaryMotion Modifying Trajectory}
\verb|Verbs| serve as the primary motion, and the \verb|PrimaryMotion| class owns the entity's trajectory. Verbs can be \verb|Solo|,  \verb|Relational|, \verb|PathMotion|, or \verb|PinMotion|. \verb|Solo| verbs are actions like \verb|struggle|, \verb|walk|, and \verb|idle|. \verb|Relational| verbs are actions that occur in between characters such as \verb|follow|, \verb|chase|, and \verb|avoid|. \verb|Solo| verbs have helper force functions like \verb|impulse| and \verb|spring|, and \verb|Relational| verbs have helper force functions like \verb|bubble|, \verb|lookAt|, \verb|attract|, \verb|repel|, \verb|collision|, \verb|mirror|, and \verb|match|. \verb|PinMotion| allows a verb to happen in-place. \verb|PathMotion| allows entities to follow along \verb|Paths|. Paths can be generated programmatically (a figure-eight orbit) or take in custom drawn paths. This allows \textit{fog} to flexibly support both complex trajectories (arcs, spirals, polylines) as well as bind to paths users freehand sketch. 

\subsubsection{Adverbs as Secondary Motion Modifying Manner}

\verb|Adverbs| are a class that can add a layer of secondary motion on top of the primary motion. At the setup of each state, an \verb|Adverb| as \verb|SecondaryMotion| can change certain qualities and channels of the motion (easing, positional offsets, energy curve). During each frame of the state, they can also directly apply additional motion within a bounded delta of the original \verb|PrimaryMotion|. These can be effects like a sway, a wiggle, jitter, hesitation, scaling, and so on. For example in Fig.~\ref{fig:teaser}, for \verb|aggressively(chase(B))|, the adverb \verb|aggressively| can provide a sharp jitter through positional offsets, to compose with a high speed \verb|chase| verb, while the \verb|chase| updates the entity towards a target each frame.


\subsubsection{Emotions as Secondary Motion Modifying Feeling}

\verb|Emotions| also extend \verb|SecondaryMotion| class to change the feel of the motion. For example, an entity can \textit{excitedly} or \textit{angrily} \verb|chase| another entity. \verb|Emotion| functions can modulate the energy on the motion performance (inspired by \cite{Strassmann:1991:DesktopTheater}) and also implement \verb|update()|. These layers would also add a bounded per-frame delta. For example, an \verb|update()| for \verb|Fear| could add a tremor on top of the entity's \verb|PrimaryMotion|. 


In this way, \textit{fog}'s class abstractions allow users to create \verb|Verbs|, \verb|Emotions|, and \verb|Adverbs| that map and combine as words (e.g. anxiously(chase)), while their functions compose as code.

\begin{table}[H]

    \caption{Types of Function Compositions in \textit{fog} Animation Framework}
    \Description{Headers: Function Compositions, Examples. Five rows of different possible function compositions emulating how language composes: Relational Verbs - Entity A chases entity B, Adverb composing with verbs - entity A hesitantly paces, Adverb composing with a relational verb - entity A hesitantly follows an entity, Path Verb composing with a Path object - entity A zigzags along a path, or entity A follows a user-drawn path, and lastly Emotion composing with a Verb - entity A angrily chases an entity.}
  \label{tab:compositions}
  \begin{tabular}{cl}
    \toprule
    Function Compositions & Example  \\
    \midrule
    Relational Verb & \verb|chase(B)| \\
    Adverb + Verb &  \verb|hesitantly(pace)|  \\
    Adverb(Relational Verb) & \verb|hesitantly(follow(B))|  \\
    PathVerb + Path & \verb|zigzag(Path)|, \verb|follow(drawPath)|  \\
    Emotion(Verb) & \verb|angrily(chase(B))| \\
    
    \\
  \bottomrule
\end{tabular}

\end{table}

\subsection{\textit{fog} in Interface form}

\begin{figure*}[t]
    \centering
    \includegraphics[width=0.9\textwidth]{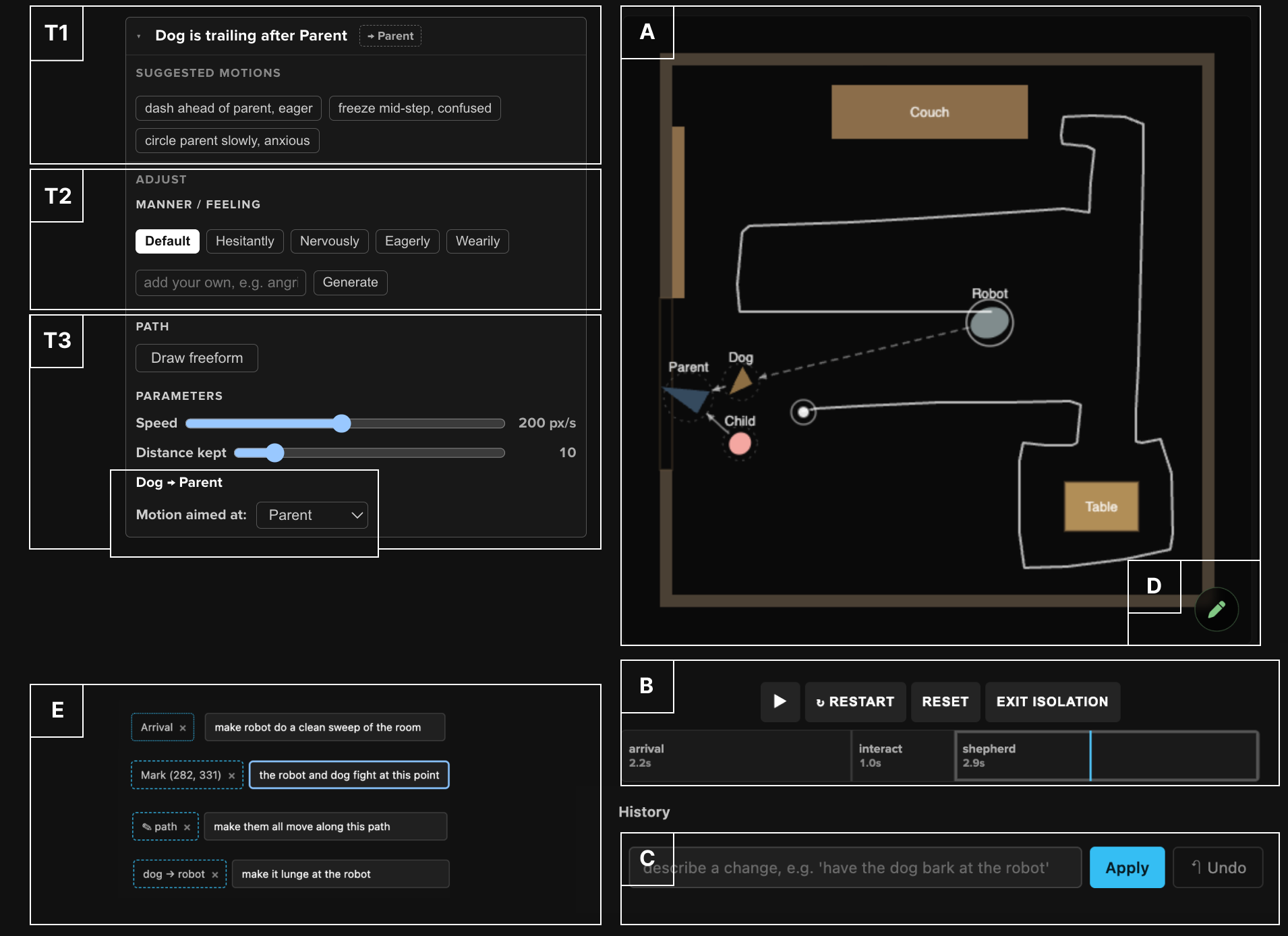}
    \caption{\textit{fog} in interface form showing a home robot animation scene (A, B) being edited (C). T1-T3 show the tiered editing. T1 is the primary layer, T2 is the secondary layer, and T3 splits out parameters from the primary layer. E shows ways we spatially and temporally localize the animation edits.}
    \Description{fog in interface form. A shows the home robot animation scene. It has a robot, dog, parent, and child. Arrows show relationships between the entities. A hand drawn path is placed on the screen. There are callouts showing the different tiers of editing (T1-T3) There is a section E which describes the different types of spatial context. Section B shows the timeline and playback controls. Section C shows the prompt. Section D marks out the pencil users can use to draw.}
    \label{fig:fog-ui1}
\end{figure*}

\textit{fog} helps users in interactive form explore these different layers of motion for Heider-Simmel type animation. It also provides a ladder of controls that support both storytelling and perceptual refinement. Now we provide a walkthrough of \textit{fog} in interface form, which is shown in Fig.~\ref{fig:fog-ui1}.

\subsubsection{Globals: Canvas, Timeline, Prompt} \textit{fog} has a Main Canvas, a Timeline, and a dedicated State Editing Panel. The animation plays out on the canvas (Fig.~\ref{fig:fog-ui1}-A). The Timeline (Fig.~\ref{fig:fog-ui1}-B) helps users visualize the events of the animation in terms of \verb|States|. Users use the playback controls to scrub through and rewatch the animation. Underneath the timeline, there is a global prompt box (Fig.~\ref{fig:fog-ui1}-C).

\subsubsection{State Editing Panel} Clicking on a \verb|state| opened up an isolation mode for dedicated editing (Fig.~\ref{fig:fog-ui1}-T1-T3). At the top of each \verb|state|, there is a description of what the entity is doing in the state. For example, in Fig.~\ref{fig:fog-ui1}-T1, we see the description ``Dog is trailing after Parent" in the first state. T1 refers to Tier 1, the first rung of \textit{fog}'s animation editing ladder, which covers natural language edits to change the \verb|PrimaryMotion| layer. In T1, \textit{fog} also provides suggested motions as chips, so users can take system-generated ideas. For example, the dog could ``freeze mid-step, confused" or ``dash ahead of the parent slowly, anxious".  

The next tier is T2, which is where the user can change \verb|SecondaryMotion| layer. They could keep the main \verb|PrimaryMotion| (T1) but apply different \verb|SecondaryMotion| layers like \verb|Hesitantly|, \verb|Nervously|, \verb|Eagerly|, or \verb|Wearily|. This primary-secondary split allows user to keep the trajectory the same, but modify the performance of the motion. \textit{fog} would provide default \verb|SecondaryMotion| presets (e.g. \verb|hesitantly|, \verb|nervously|) but it would also always generate new \verb|SecondaryMotion| functions tailored to the current animation each time a prompt was fulfilled. There was also a secondary prompt area where users could generate new \verb|SecondaryMotion| functions on-the-fly. For example, a user could type in \verb|excitedly| and have \verb|excitedly| be generated as a new \verb|SecondaryMotion| instance.

The last tier was Tier 3, where users had the most direct forms of control. For example, if the user wanted to change the path the animation was taking, they could draw a path that would be deterministically bound to the entity. If the user wanted to change the speed, they had sliders. In this layer, we also surfaced variables from the generated code. For example, in the \verb|RelationalVerb| \verb|follow|, there is a \verb|distanceKept| specific to that \verb|Verb| motion that users could manipulate.

\subsubsection{Generating with Prompts that Localize Timeline and Canvas Edits}
Every time a user entered a new prompt, the code representing the animation is regenerated. This code holds the \verb|Scene|, its \verb|States|, \verb|Entities|, \verb|PrimaryMotion|, and \verb|SecondaryMotion| layers attached to each entity. For example, in an animation involving a robot, a dog, parent, and a child shown in Fig.~\ref{fig:fog-ui1}, they could pass in edits like ``have the dog bark at the robot". The verb \verb|bark| would be generated with the \verb|RelationalVerb| type for \verb|dog|, and robot would be the \verb|target|. The \verb|bark| would become the new \verb|PrimaryMotion| set on the dog. Alongside the generation of the new verb, \textit{fog} also generates \verb|SecondaryMotion| functions (e.g. \verb|aggressively(bark)| and \verb|excitedly(bark)|), so users can modify the manner and feeling on \verb|bark|.

\textit{fog} also gives users simple controls to localize what part of the animation they are trying to edit. The different forms of spatial or temporal context for prompts are shown in Fig.~\ref{fig:fog-ui1}-E: targeted state, marks, arrows, and paths. For example, if a user clicks on the state named ``Arrival" in the Timeline, a chip with the word ``Arrival" appears by the prompt. In the backend of the system, this appends context to the prompt stating that it should only return a state-scoped edit, rather than regenerating the entire animation scene. If user clicks on the Pencil tool (Fig.~\ref{fig:fog-ui1}-D) and makes a on the canvas, a chip with ``Mark: (x,y)" appears by the prompt box. This appends the mark location within the prompt, so a user can point on the canvas and say ``make the entities meet here ($\odot$)". If a user draws a path with the Pencil tool, this appends a down-sampled, 48-point version of the path within the prompt. This way users can pass paths into prompts and say ``make the entities fight along this path ($\sim$)". If the user draws an arrow with the Pencil tool, this communicates to \textit{fog} that A relates to B and that a \verb|RelationalVerb| should be generated. This allows the user to express interactions between entities like ``make it lunge at the robot". Examples of marks, arrows, and paths are shown on the canvas in Fig.~\ref{fig:fog-ui1}-A. This allowed us to satisfy DG4 to create ways for users to provide spatial and temporal context that localizes the prompt. Since these prompt pipelines were long, we utilized prompt caching. The prompts that backed our user interface are detailed in Supplementary Material.



\section{Evaluations}

We evaluated \textit{fog} through two evaluations to understand its claims and capabilities as a framework. The first evaluation was a technical evaluation benchmarking how capably LLMs could generate motion functions and function compositions using the code abstractions for \verb|Verb|, \verb|Adverb|, \verb|Emotion|, \verb|PrimaryMotion|, and \verb|SecondaryMotion|. The second evaluation was a user study to understand how novices to animation could utilize this framework to create Heider-Simmel type animations.

\subsection{Technical Evaluation of fog Framework}

The first evaluation addressed the following research question: RQ1) How do the code abstractions in \textit{fog}'s framework change the quality and expressive range of a base LLM at generating animation? RQ2) Is there any difference between generating one function that holds the entire motion or two functions that compose as separate layers? 

We hypothesized: H1) \textit{fog}'s code abstractions would improve the quality of LLM-generated animations, because they suggest channels the LLM could manipulate like easing, offsets, and acceleration. We also hypothesized H2: motion functions could be generated separately and composed together, just as manner and feeling words have their own independent meaning from verbs.




\subsection{Technical Evaluation Methodology}
To address RQ1 and RQ2, we looked at four conditions of LLM animation generation that imposed different levels of structure. In each condition, the LLM was tasked to generate animation on a circle for a stimulus word or phrase like ``chase" or ``hesitantly chase".

\begin{itemize}
        \item R1) \textbf{Unstructured}. LLM generates raw code document to express stimulus word in terms of animation. There are no structure constraints on the code.
        \item R2) \textbf{Engine}. LLM generates one function called \verb|step()| to express the stimulus word in terms of animation. \verb|step| is rendered by fog's animation engine without base secondary layer effects (squash-and-stretch and motion trail). 
        \item R3) \textbf{Engine + Layer}. LLM generates one function called \verb|step()| to express the stimulus word in terms of animation. \verb|step| is rendered by fog's animation engine with base secondary layer effects (squash-and-stretch and motion trail). 
        \item R4) \textbf{Structured}. LLM generates two typed functions (Verb, Adverb, Emotion, Emotion(Verb), Adverb(Verb) and composes them as two distinct layers. The functions implement expected methods like \verb|update()| and other methods bespoke to each class. At composition time, functions compose on a discrete set of channels (easing, position offsets, timing, acceleration, speed). State is separated across two functions.
\end{itemize}

We then generated a variety of samples for each condition using stimulus words and phrases for the following motion dimensions: \verb|Verb|, \verb|Adverb|,  \verb|Emotion|, \verb|Adverb(Verb)|, and \verb|Emotion(Verb)|. These stimulus words were chosen to balance for variety and coverage. For adverbs, we chose 12 split across speed and manner. For verbs, we chose 32 words balanced across a 2x2 split for solo vs. relational motion and path vs. non-path motion. For emotions, we chose 12 emotions balanced for valence. The exact words are shown in Table ~\ref{stimulus-words}. Then for each condition R1-R4, we generated animations expressing these sets of stimulus words. We tasked \verb|Claude-Fable-5| to implement these stimulus words with the following parameters: maxTokens at 16,000 tokens, adaptive thinking, and effort set to ``high".  Our prompt pipelines are detailed in Supplementary Material.

\begin{table}[htbp]
\centering
\caption{Stimulus Words for Technical Evaluation of Motion Function Generation}
\label{tab:lexicon}
\vspace{0.5em}
\renewcommand{\arraystretch}{1.25}
\begin{tabularx}{\textwidth}{l l X}
\toprule
\textbf{Category} & \textbf{Subcategory} & \textbf{Items} \\
\midrule
\textbf{Adverbs} (12)  
  & Speed & slowly, quickly, suddenly \\
  & Drive & urgently, eagerly, forcefully \\
  & Caution & hesitantly, nervously, lazily \\
  & Manner & erratically, gently, playfully \\
\midrule
\textbf{Verbs} (32)  
  & Solo non-path & idle, tremble, struggle, throb, flinch, shrink, wander, bounce \\
  & Solo path & walk, spiral, zigzag, pace, glide, dart, swoop, lunge \\
  & Relational non-path & follow, avoid, flee, cling, guard, watch, repel, hug \\
  & Relational path & chase, orbit, escort, mirror, weave, intercept, flank, leapfrog \\
\midrule
\textbf{Emotions} (12)  
  & Negative & anger, confusion, disgust, fear, boredom, sadness \\
  & Positive & ecstatic, happy, surprise, awe, relaxed, tenderness \\
\bottomrule
\label{stimulus-words}
\Description{All stimulus words in columns organized by Category, Subcategory, and Items. Adverbs: slowly, quickly, suddenly, urgently, eagerly, forcefully, hesitantly, nervously, lazily, erratically, gently, playfully. Verbs: idle, tremble, struggle, throb, flinch, shrink, wander, bounce, walk, spiral, zigzag, pace, glide, dart, swoop, lunge, follow, avoid, flee, cling, guard, watch, repel, hug, chase, orbit, escort, mirror, weave, intercept, flank, leapfrog. Emotions: anger, confusion, disgust, fear, boredom, sadness, ecstatic, happy, surprise, awe, relaxed, tenderness. }
\end{tabularx}
\end{table}

\begin{figure}[h]
        \centering
         \begin{tcolorbox}[
        colframe=gray, 
        colback=white,     
        arc=4pt,           
        boxrule=1pt,       
        halign=center,
        left=2pt, right=2pt, top=2pt, bottom=2pt 
        ]
        
        \includegraphics[width=\linewidth]{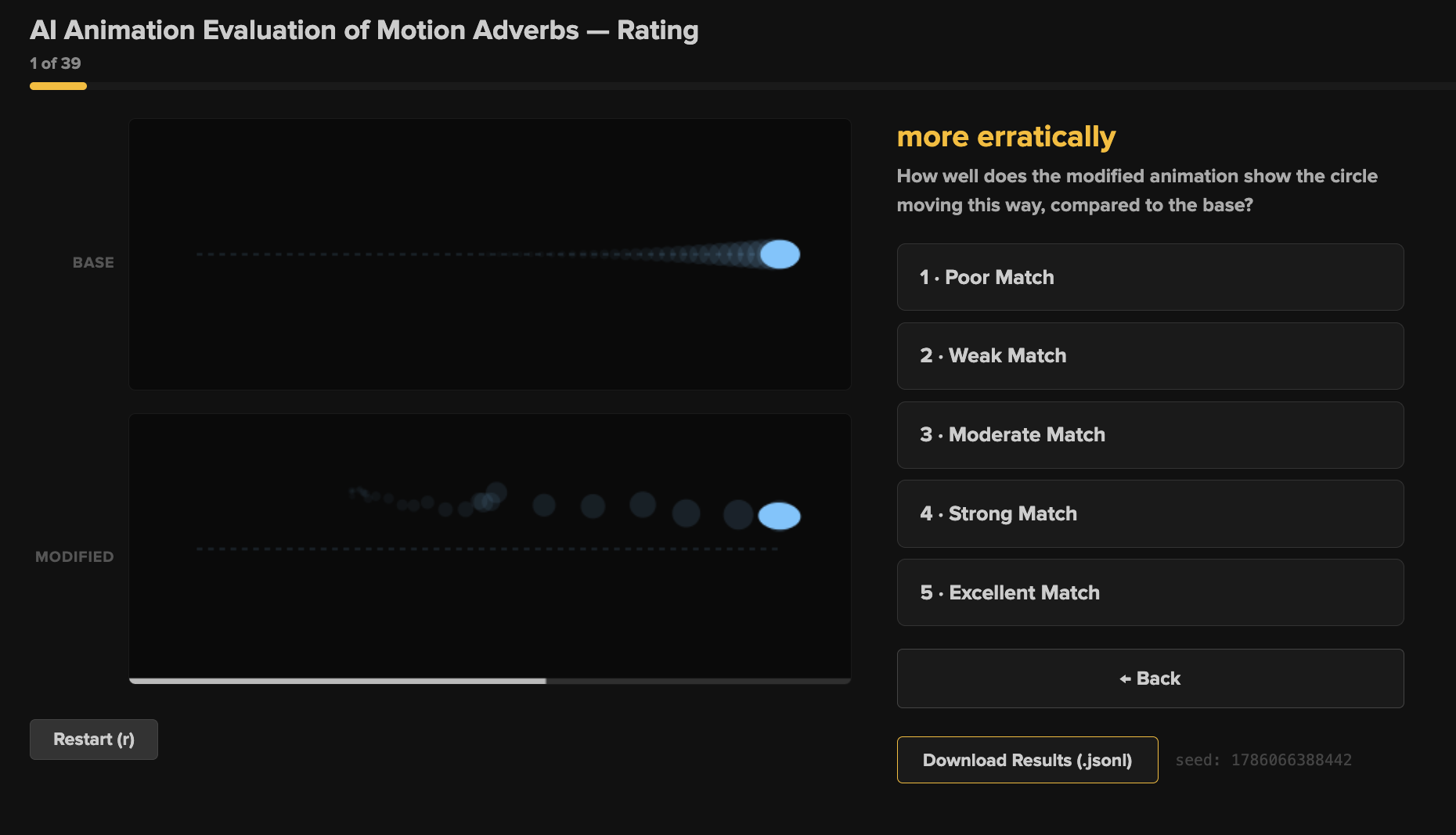} 
        \end{tcolorbox}
        \caption{\textit{fog}'s technical  evaluation had crowdworkers rate generated animations on a 1-5 scale using a page like the following.}
        \Description{Example of the viewer shown to raters for User Study 1. The rated animations for Verb, Adverb, Emotion, Adverb x Verb, and Emotion x Verb. This image illustrates the adverb task, where there is a base animation and a modified animation (two static frames where the motion trails are evident is shown). On the left side there is a 1-5 scale reading: 1 poor match, 2 weak match, 3 moderate match, 4 strong match, and 5 excellent match.
        }
        \label{fog-rating-adverbs}
        
    \end{figure}

Once the source code for the animations were generated, we created an experimental apparatus page where the animations from all four conditions would be loaded in randomized order, so that crowdworkers could rate them. For \verb|Emotion| and \verb|Verb|, we rendered the animation on a canvas and asked crowdworkers, on a scale of 1-5, how well does the animation match the given \verb|Emotion| or \verb|Verb|? For \verb|Adverbs|, we provided a base animation and an \verb|Adverb|-modified animation and asked, ``How well does the modified animation show the circle moving more ADVERB compared to the base?" See Fig.~\ref{fog-rating-adverbs}) for an example of what a rater saw when evaluating \verb|Adverbs|. For \verb|Relational| verbs, a second red circle was also present so the entity could express the motion on a target. 

The study was conducted on CloudResearch Connect. Each crowdworker rated all the stimulus words for the dimension. The median time for this experiment per task was 6.37 min for \verb|Adverbs|, 6.49 min for \verb|Emotion|, and 6.40 min for \verb|Verbs|. These durations included the time taken to read instructions and take two practice trials. During the task, we implemented attention checks to distinguish between bots and humans. The attention check was to pick a number that briefly flashed on the canvas. The motion was fast so that it would be hard to capture and send to a VLM. It would also be harder for a VLM to respond faster than a human, especially across three attention checks. Furthermore, we did a posthoc inspection of crowdworkers to flag for 1) outlier time taken 2) low accuracy below chance, and 3) position bias (always clicking the same answer). Across all five conditions, we excluded ratings from participants who failed attention checks and the post-hoc analysis. Participants were paid \$12-15/hr wage across tasks, with shorter rates for the tasks reviewing less samples (e.g. \verb|Adverbs| or \verb|Emotions| conditions).



\subsubsection{Technical Evaluation Results for Adverbs, Verbs, and Emotion}

    \begin{figure}
        \centering
\includegraphics[width=\textwidth]{./figures/fog_ladder_v2.pdf} 

        \caption{Results of ladder evaluation for motion dimensions (Verb, Adverb, Emotion) and function compositions (Adverb(Verb) and Emotion(Verb) across the rungs R1-R4, which increase amount of structure imposed on LLM. The bars represent the mean aggregated across raters, with 95\% CI. Brackets going across rungs show significance $p \leq 0.05.$ }
        \label{fog-ladder}
    \end{figure}



For the \verb|Adverbs| dimension, we report R1 (Unstructured): M = 2.96, $\sigma = 0.71$; R2 (Engine): M = 3.88, $\sigma$ = 0.86 ; R3 (Engine + Layer): M = 3.90, $\sigma$ = 0.95; R4 (Structured): M= 3.14, $\sigma$ = 0.70. We applied the Friedman's chi-squared test (non-parametric alternative to repeated-measures ANOVA, does not assume normality). We found the difference between the means to be significantly different ($\chi^2(3)=$19.08, df = 3, $p$ = 0.0003). This allowed us to proceed to test the significance between rungs using four Wilcoxon signed-rank tests with Holm-Bonferroni correction. We tested between R1-R2, R2-R3, R3-R4, and R1-R4, finding significance between R1-R2 ($M_{\text{diff}} = 0.91$, Holm-adjusted $p = 0.0078 \leq 0.05$) and R3-R4 ($M_{\text{diff}} -0.76,  $ Holm-adjusted $p = 0.016 \leq 0.05)$ This means that in the Adverbs condition, R2 and R3 were significantly higher than R1. The ratings went up by about one full point. However, R4 was significantly lower than R3. The human-interpretable version of these numbers is that \textit{fog}'s animation engine structure helped in the case of \verb|Adverbs|, but the two-function setup of the Structured condition did not meaningfully improve animation quality over the Unstructured LLM condition. This is illustrated in the second barchart (top right) of Fig.~\ref{fog-ladder}.

For \verb|Verbs|, we report R1 (Unstructured): M = 4.05, $\sigma = 0.58$; R2 (Engine): M = 3.50, $\sigma$ = 0.32 ; R3 (Engine + Layer): M = 3.61, $\sigma$ = 0.56; R4 (Structured): M= 3.20, $\sigma$ = 0.35. Again, we applied a Friedman's test. We found the means across R1 and R4 to be significantly different (Friedman chi-square = 14.21, df = 3, p = 0.003). This allowed us to proceed to test the significance between rungs again. We did four Wilcoxon signed-rank tests with Holm-Bonferroni correction looking between rungs: R1-R2, R2-R3, R3-R4, and R1-R4. We found significance between R1-R2 ($M_{\text{diff}} = -0.55$, Holm-adjusted $p = .047 \leq 0.05$) and R1-R4 ($M_{\text{diff}}= -0.85$, , Holm-adjusted $p = .016 \leq 0.05)$ This means that in the Verbs condition, R1 was significantly higher than R4. The structure of the code class reduced the quality of \verb|Verbs| generated compared to the unstructured condition. This is illustrated in the first barchart (top left) of Fig.~\ref{fog-ladder}. 

For the \verb|Emotion| dimension, we report R1 (Structured): M = 3.56, $\sigma = 0.58$; R2 (Engine): M = 3.30, $\sigma$ = 0.65 ; R3 (Engine + Layer): M = 3.32, $\sigma$ = 0.71; R4 (Structured): M= 3.51, $\sigma$ = 0.75. We applied the Friedman's chi-squared test again. We found the difference between the aggregated means across R1, R2, R3, and R4 to be not significantly different ($\chi^2$(3)=7.483, $p$ = 0.06). This is illustrated in the second barchart (middle row, left) of Fig.~\ref{fog-ladder}. In this case, the added structure of the \verb|Emotion| class did not make a meaningful difference.

\subsubsection{Function Composition Rating Methodology}
After the first three dimensions were evaluated, we picked the eight words from each dimension (\verb|Verb|, \verb|Adverb|, \verb|Emotion|) and crossed them to look at two types of function compositions we wanted \textit{fog} to support: \verb|Adverb(Verb)| and \verb|Emotion(Verb)|. We used a script to generate 16 stimulus phrases (e.g. ``hesitantly chase", ``slowly wander", ``angrily spiral") that balanced across the clusters within each category. We generated function compositions for conditions R1-R4. For R1-R3, the generation procedure was essentially the same, except that instead of a stimulus word as a prompt, the stimulus phrase was passed in (e.g. ``generate \verb|step()| for the following phrase: \textit{hesitantly chase}". For R4, we generated two functions (\verb|Adverb|, \verb|Verb|) or (\verb|Emotion|, \verb|Verb|) and composed them on the page when the animation was run. This composition composed on a discrete set of channels (acceleration, speed, easing, timing, pauses). To evaluate these function compositions, again we used the same rating page structure and had crowdworkers rate the animations on a scale of 1-5.

    


\subsubsection{Function Composition Rating Results.}

For the \verb|Adverb(Verb)| dimension, we report R1 (Unstructured): M = 3.86, $\sigma = 0.29$; R2 (Engine): M = 3.50, $\sigma$ = 0.59 ; R3 (Engine + Layer): M = 3.52, $\sigma$ = 0.36; R4 (Structured): M= 2.84, $\sigma$ = 0.32. We applied the Friedman's chi-squared test, finding the difference between the aggregated means across R1, R2, R3, and R4 to be significantly different ($\chi^2$(3)=18.77, $p$ = 0.0003).
In the Adverb(Verb) condition, R1 was significantly higher than R4, and R3 was significantly higher than R4. This showed that composing Adverbs and Verbs on five contracted channels (easing, timing windows, acceleration, speed, pauses) using two functions was worse than generating one monolithic function in R3. Even though at generation time, we generated \verb|Adverbs| and \verb|Verbs| with prompt context that they were layers meant to compose, the channel-specific composition created a ceiling in expressive range (they only had five channels to use), and ratings went down for R4. This is illustrated in the second barchart (middle row, right) of Fig.~\ref{fog-ladder}.

For the \verb|Emotion(Verb)| dimension, we report R1 (Unstructured): M = 3.81, $\sigma = 0.86$; R2 (Engine): M = 3.32, $\sigma$ = 0.78 ; R3 (Engine + Layer): M = 3.46, $\sigma$ = 0.82; R4 (Structured): M= 3.01, $\sigma$ = 0.73. We applied the Friedman's chi-squared test, finding the difference between the aggregated means across R1, R2, R3, and R4 to be significantly different  ($\chi^2$(3)=13.91, $p$ = 0.0003). This means that in the \verb|Emotion(Verb)| condition, the difference between ratings between R1 and R4 was again significant. For the \verb|Emotion(Verb)| condition, the ratings significantly decreased from R1 to R2 and from R3 to R4. R1 was also significantly higher than R4. This showed that similar to the \verb|Adverb(Verb)| case, composing separate functions for Emotion and Verbs was worse than generating one monolithic function in R3. This is illustrated in the second barchart (bottom row, right) of Fig.~\ref{fog-ladder}.

\subsubsection{Ladder Evaluation Conclusion.}
Overall, this ladder evaluation showed us that there is a trade-off between structure and expressive range in code generation. Structure is compelling because it lends to intermediate representations that we can use downstream in the user interface. However, the more we imposed structure in terms of class abstractions, the more we reduced the expressive range or quality of animations the LLM can achieve. The biggest reduction came from R4 in compositions, when we forced functions to compose on a specific set of channels. However, we cannot get around imposing structure if we want to create an animation system powered by LLM code generation. 

\textbf{Our main insight from this study is that separating state across two functions and restricting composition to specific channels did not help, but keeping state in one monolithic function did.} This finding led us to shift our editor to implement a hybrid approach between R3 and R4. We wanted to keep the separability of layers that R4 allowed for, because that provided a place for interactive control -- a primary-secondary split. However, we recognized that the channel-specific composition of R4 was too limiting. We adapted to an approach where each prompt would generate or update the entity's \verb|PrimaryMotion| as a monolithic function, with separate \verb|SecondaryMotion| functions. This gave us swappability and composition with layers while keeping the LLM less constrained in what channels it could use to express motion. We used this approach going forward in Study 2, our user evaluation.

\section{User Study}
Our user study addressed the following research question: RQ3) To what extent could fog's tiered editing support users in terms of control and workflow iteration? Our hypothesis was that the ladder from semantic edits to precision controls for speed and parameter editing as well as the spatial context prompting (marks, paths, arrows) would give users more control.

\subsection{User Study Methodology}

We recruited participants on Upwork by putting up a job posting for user testers of an AI animation prototype. All had experience in usability testing and an interest in animation. Four participants had exposure to animation through game development generative AI content creation tools. Participants were paid \$20/hr of their time. They were sent an information sheet and consent form beforehand. This methodology is covered by a relevant IRB protocol.

During the study, participants were given an introduction to \textit{fog}’s interface through a tutorial and walkthrough of each feature. They were then told that their experimental task was to animate two Heider-Simmel stories with shapes, starting from a base animation. The first base animation was a home robot animation involving a robot, dog, child, and parent. The second base animation involved six characters and showed an outsider being rejected by a group with two bystanders on the side. Participants first animated one scene using the full \textit{fog} editor and another scene using a baseline editor. The baseline editor was a prompt-only version of the full \textit{fog} editor. It did not have the T1-T3 controls or motion editing panel. It just provided users with a prompt textbox and a timeline to visualize the animation events. 

\begin{figure}[h]
        \centering
         \begin{tcolorbox}[
        colframe=gray, 
        colback=white,     
        arc=4pt,           
        boxrule=1pt,       
        halign=center,
        left=2pt, right=2pt, top=2pt, bottom=2pt 
        ]
        \includegraphics[width=\linewidth]{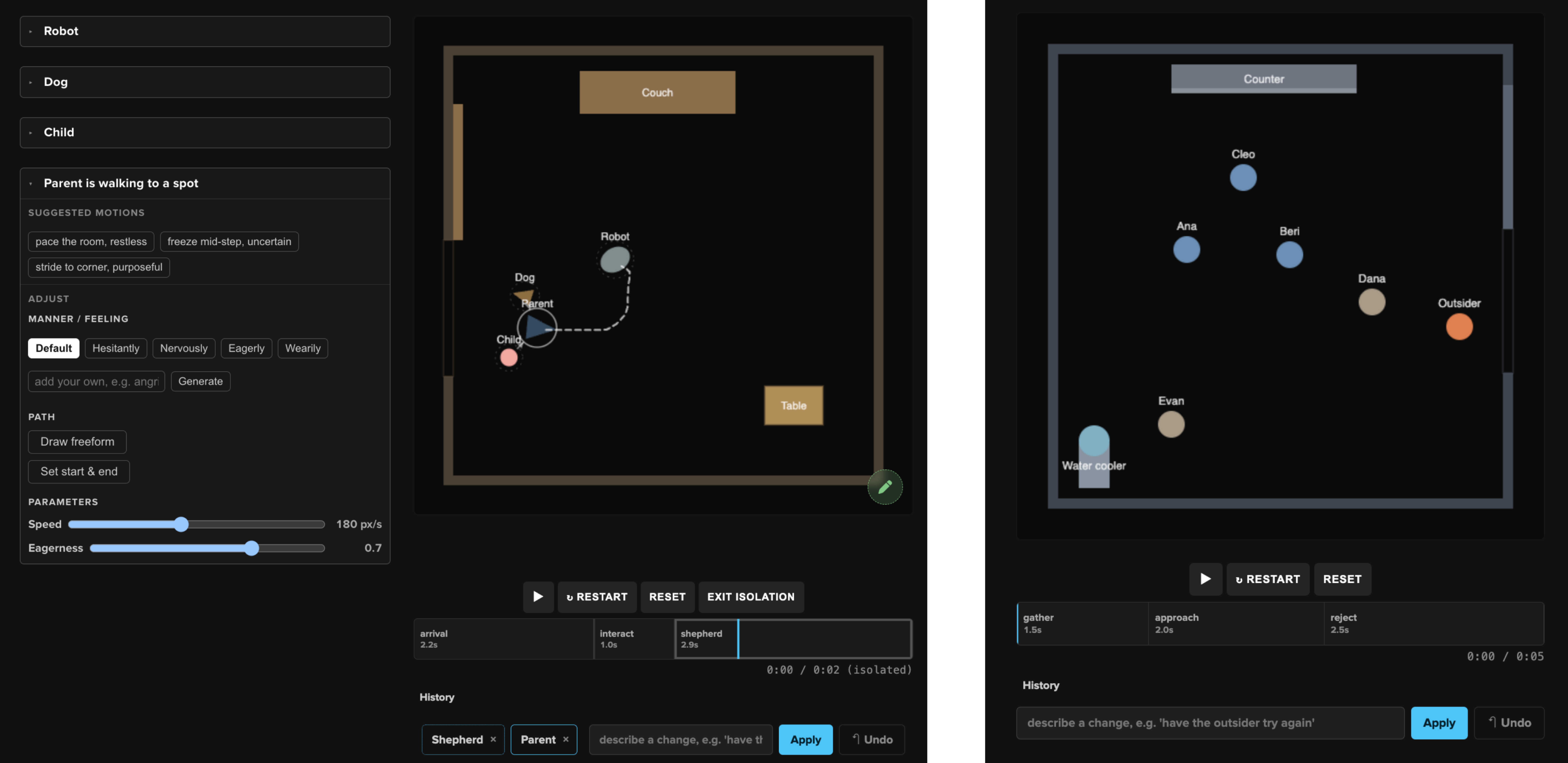} 
        \end{tcolorbox}
        \caption{\textit{fog}'s full editor (left) versus baseline (right) on two scenes. These scenes were base animations for our user study that every participant edited.}
        \Description{We show our editor on the left with its three tiers of editing. The motion refinement editor is expanded on the parent entity. Users can see suggested motions, how to adjust manner / feeling in terms of secondary layers like hesitantly, nervously, eagerly, wearily. They can draw paths using the draw freeform button and edit parameters. On the right, we show the other scene, a break room scene, where the timeline is shown but there is only a prompt box.  }
        \label{fig:baseline-scenes}
    \end{figure}

Prior to beginning the animation editing experience, we had them consider the base animation and write down an objective for how they would like to edit the animation. This was to help them try to actualize a concept they had in mind and do targeted editing, as opposed to being completely unconstrained during the task. Then, participants had 20 minutes to edit the base animation using either \textit{fog} or the baseline. The baseline and animation scenes are pictured in Fig.~\ref{fig:baseline-scenes}. To minimize for learning effects, we randomized the order of the prototype and scenario type across participants. Between each task, we had participants fill out a survey with Likert-scale questions about control and workflow iteration. Additionally, we instrumented the prototype to log interaction traces.

\subsection{User Study Results}

Overall, participants described both versions of \textit{fog} to be enjoyable and performant experiences. Though they all started from the same base scenes, the novices took the animation in many different directions. They declared animation objectives like: make ``the parent and the child start arguing, so the dog hides behind the couch and the child hides behind the table" or ``have the outsider bring the group water so they can accept the outsider", inverting the starting social dynamic of rejection. 

We first report the more plain findings -- what was unanimously liked and disliked across participants. First, participants also found the features of the tool intuitive and easy to grasp. They understood the motion layers of primary and secondary motion and could map them to socially meaningful trajectories and interactions. They liked that they could pick up a pen tool to shape the trajectory directly and enjoyed the feeling of binding paths to actions. They also liked how creating \verb|SecondaryMotion| like a little jitter occurring periodically between two characters represented chatting, or some slight rotations on a triangle Parent represented yelling. The timeline was also appreciated as a source of truth for the animation. P6 described how seeing the different states laid out in the timeline allowed them to maximize accuracy when they chose to localize their edits on states.

Participants unanimously did not like the latency that came with the editing experience in both \textit{fog} and the baseline. Each edit could take up to 20 to 90 seconds, with the longest one stretching to 2 minutes for one participant. The reason was that if the edit was very complex -- for example if the prompt was very long and touched every state -- the entire scene was rewritten, and regenerating each token produced a consistent wait-time that stretched over time. This led participants to try different strategies in \textit{fog} and the baseline. Whereas in the baseline people began to handoff more and more work to the AI, binning multiple changes in one text prompt, in \textit{fog} some participants deliberately tried working more step-by-step.

\subsubsection{Dimensions of Control}
In our exit survey, we asked each participant which interface gave them more control over their animation editing experience. Eight of 12 participants said \textit{fog} answered in the survey that the full \textit{fog} interface them more control. Many participants verbally expressed frustration trying to manipulate timing and spacing in the baseline condition. They appreciated that there were methods to address these painpoints in \textit{fog}'s editor. For example, P4 said ``\textit{I find the second version [fog] a lot easier to control. I like how the drawing was fast, and it could be very precise. The robot trajectory, for example. I would imagine if I just used the prompt-based version, it would take a few iterations to get it the way I wanted. Whereas when I drew the trajectory  I wanted it to follow, there was such little processing time -- I was delighted. And in terms of [editing] the speed in the first version [baseline], I thought that was actually a struggle. I just generally found the second version [fog] to be much more customizable. It gave me more precise control.}" P11 expressed a similar sentiment, but stated that because there were more controls, while good, there were more places to break, ``\textit{The second one definitely gave me more control, which made it easier to mess up.}" We now dive into the specifics of each dimension of control \textit{fog} looked at.

\subsubsection{Spatial Control}
There was no significant difference between users in terms of spatial control across \textit{fog} and the baseline. The median rating for both fog and the baseline was 6/7 -- both scores saturated at the high end of the Likert scores. While we could have been statistically underpowered to find a difference in Likert scores, we had thought providing the pen tool would provide significantly more control. Instead, we found that for some participants, the baseline was sufficient for control so long as the participants could \textit{name} where they wanted things to go. It was enough to be able to tell the robot to go to the corner, or the child to go to the table.

\begin{figure}[h]
        \centering
         \begin{tcolorbox}[
        colframe=gray, 
        colback=white,     
        arc=4pt,           
        boxrule=1pt,       
        halign=center,
        left=2pt, right=2pt, top=2pt, bottom=2pt 
        ]
        \includegraphics[width=\linewidth]{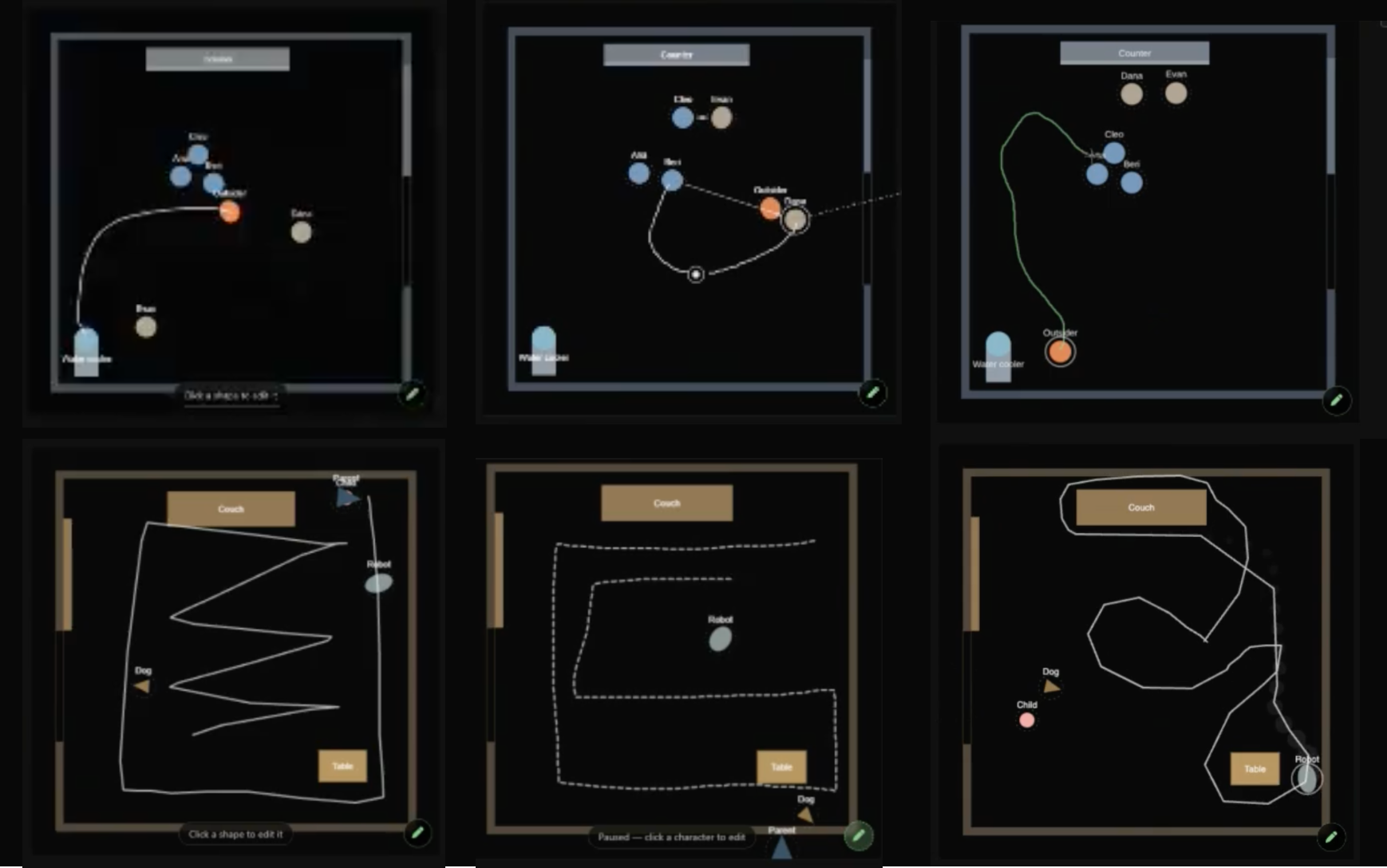} 
        \end{tcolorbox}
        \caption{Examples marks participants drew to point at places on the canvas, paths participants drew, and arrows participants created to express relationships. These marks were translated into data and context within the prompts for AI to implement.}
        \Description{Examples of paths that users drawn in a 2x3 format. Users drew curved, arching paths, zigzagging paths, arrows to mark out relationships, and they made marks for where characters should meet. The richer context of these spatial marks is detailed out in the paper. }
        \label{fig:paths-marks}
    \end{figure}

However, qualitatively, we heard from many participants that in the baseline, one thing AI struggled to get right was spacing. P8 described how when they would prompt for the entity to go to the water cooler in the scene, the AI edit would drop the entity off at a point that was slightly off from where they wanted to be, or it would have the entity collide with another entity on its way there. It was hard to make corrective spatial edits in the baseline condition; describing how a position should change or how there should be obstacle avoidance and line of sight is difficult to describe in words. On the other hand, in \textit{fog}'s editor, participants drew marks and paths with the pen tool as a way to handle error recovery. They also used marks to point on the canvas and drew paths that could characterize the entities (e.g. an angular path for a robot, Fig.~\ref{fig:paths-marks}, bottom middle). Some also used paths to stage interesting group dynamics. For example, P6 dropping a character off right in between the group and made everyone in the group recoil at the intrusion (Fig.~\ref{fig:paths-marks}, top right). 

P2 created the most complex combination of arrows, marks, and paths to stage a fight scene. Their objective was to make two teams of characters gather at a point on the canvas and then clash into each other before leaving the room. They were able to break that objective down into substeps, so that entities gathered at marks and then devolved into pairs that fought. These pairs were established through arrows and marks (Fig.~\ref{fig:paths-marks}, top middle). While they felt they had enough spatial control to accomplish their complex objective, they also felt that the added breakpoints in the motion and latency were a tradeoff. They thought sacrificing some spatial control would have been fine for a more one-shot editing experience.


\subsubsection{Speed and Timing Control}

We found no significant difference in the median rating between \textit{fog} and the baseline for control on speed (fog median: 6/7, baseline median: 6/7) and timing (fog median: 5/7, baseline median: 6/7). Most participants did use the speed slider in the \textit{fog} editor if they saw after watching the animation that something was too fast or too slow. However, most participants did not use the GUI interactions for manipulating time and duration. For example, they did not change the boundaries of the states much or trim down their durations. In both interfaces, they tended to  reach for text and prompts to edit timing:  ``\textit{delay the entrance of the entity by two seconds}", ``\textit{when the parent arrives, have it join the chase around the room}". This user behavior told us that: 1) in the context of Heider-Simmel animation, the ordering of events and synchronized actions can be coarsely defined and captured well enough by natural language, 2) novices new to a GUI would prefer to have AI handle perceptual refinement if there are even slight discoverability issues for where the controls are.

However, we did observe that at least in the case of two baseline conditions, participants could get stuck in frustrating loops editing speed and timing with just prompts. P4 described how it was challenging to tune speed down using just words. P11 found it difficult to fix timing issues when events were almost simultaneous: ``\textit{It didn’t look like a confrontation. The outsider's running away before it intervenes really}." When events became very close in timing, words and sentences lost the granularity to separate out the events.

\subsubsection{Narrative Control}

We anticipated that the baseline would be better or even at narrative control if users would be more split over narrative, numerical, and direct manipulation edits in the \textit{fog} interface. This was essentially what we found; fog had a median rating of 5/7 for narrative control while the baseline had a median rating of 6/7. Regardless of the condition, however, participants created unique moments through prompts. They directed scenes where entities brought water from the water cooler to friends (P8), entities drank water from the water cooler (P6), and entities like the dog latched onto the robot to ride it around the room (P6). Participants did not struggle to come up with these narrative moments and in large part did not need the ideation assistance from the suggested motion panel in T1. 

One thing we hypothesized was that in the baseline condition, there would be issues of preservability. Because people could be imprecise in their text prompts describing the animation, some elements of the animation could drop out over time as the regeneration kept occurring. This issue was indeed observed for some participants like P11. “\textit{Okay, now the robot is doing what I wanted to … but I don’t know why the other ones [events] dropped off. Since it removed parts of the scene, I'm trying to say that it should still do that.}” Participants would then switch away from story editing to handle error recovery. They would ask AI to figure out 1) why there were discontinuities in motion trajectory, such as why entities would jump from one place to another, 2) why characters disappeared from the scene (P7) and how to put them back, and 3) why events were dropped out over edits. This was a drawback of prompting that was apparent in both \textit{fog} and baseline.

\begin{figure}
        \centering
        \includegraphics[width=0.8\linewidth]{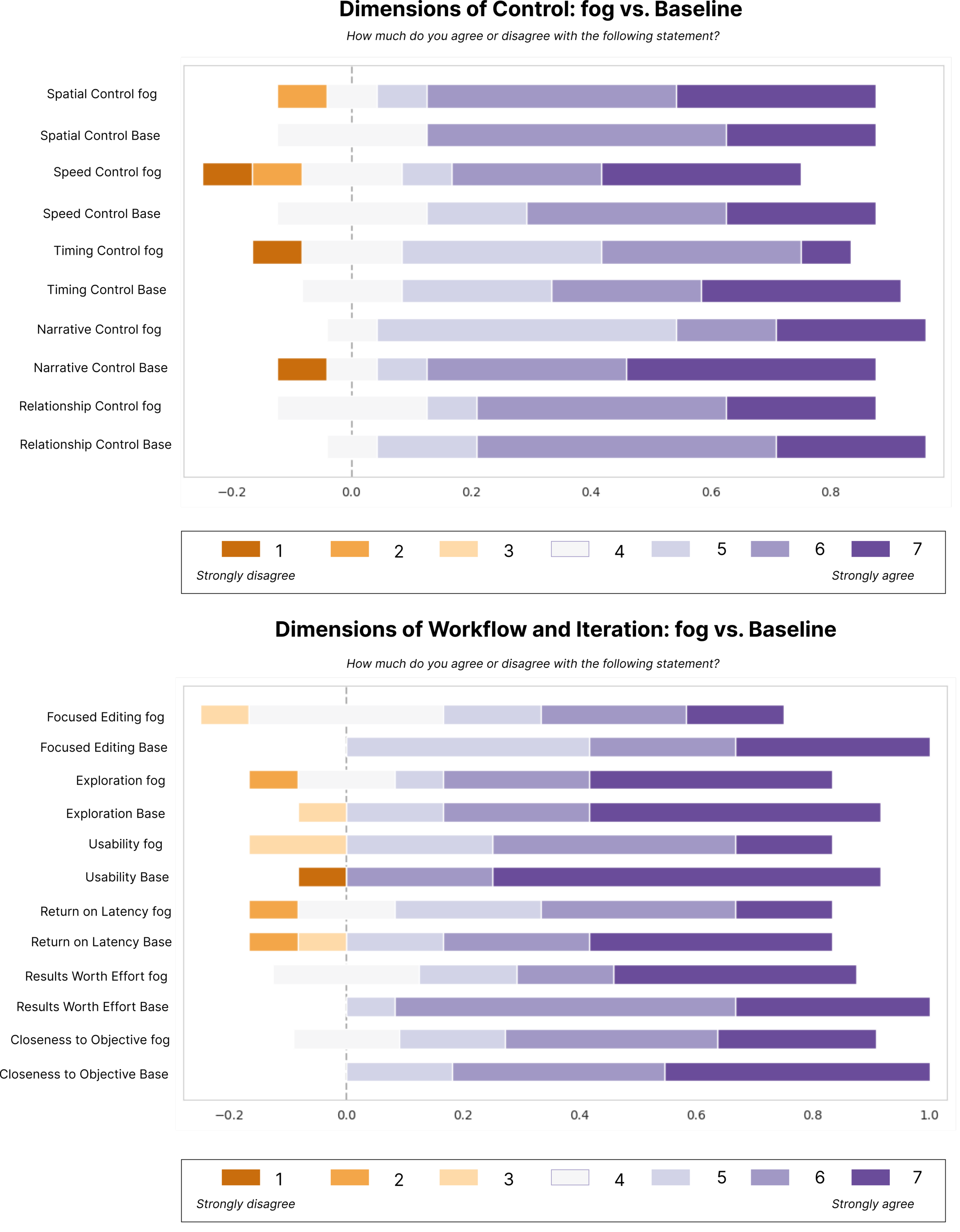} 

        \caption{Likert scale from study questionnaire about control and workflow iteration dimensions between \textit{fog} editor and baseline. The measures were not significantly different on any dimension.}
        \Description{Likert scale numbers. They show neutrality at 4 and use purple orange to designate positive and negative sentiment respectively. There are mostly positive numbers for both condition, though fog full editor gets a few more negative responses on dimensions like timing control and speed control.}
        \label{fog_workflow_control}
    \end{figure}

\subsubsection{Conclusion on Control}

Our initial hypothesis was that \textit{fog} would provide significantly more control on different dimensions, but particularly for spatial control. Quantitatively, we did not find this significance because ratings saturated at the higher end. Qualitatively, we found that the more appreciated tools were the marks and paths of the pen tool -- it gave users a way to point and gesture trajectories to AI. The other direct manipulation capabilities of manipulating speed and timing were less utilized, but still gave users deterministic knobs for control. Overall, people preferred fog as the interface they felt more control over overall (8/12 on exit survey), but the tiered controls added more breakpoints to their animation. The added cognitive load of learning a GUI and discovering its features traded off with the simpler and smoother experience of the baseline.

\subsubsection{Workflow and Iteration Measures}

We also asked in our post-task surveys about different workflow dimensions such as focused editing, exploration, return on latency, and results worth effort. Again, comparing across the median ratings with a Wilcoxon's signed rank test, we did not find any significant differences on any dimension. These measures are reported in the bottom half of Fig.~\ref{fog_workflow_control}. While we hypothesized that fog could support more focused editing or closeness to objective, on most dimensions, participants rated the baseline just slightly higher. On Focused Editing) \textit{fog} had a median rating of 5/7, while the baseline had a median rating of 6/7. On Exploration, \textit{fog} had a median rating of 6/7, while the baseline had a median rating of 6.5/7. On Usability, \textit{fog} had a median rating of 6/7, while the baseline had a median rating of 7/7. On Results Worth Effort, \textit{fog} had a median rating of 6/7, and likewise the baseline had a median rating of 6/7. On Closeness to Objective) \textit{fog} had a median rating of 6/7, and likewise the baseline median had a median rating of 6/7. Ultimately, this suggests to us that prompt baselines powered by state-of-the-art models are hard to beat -- they provide a smooth, frictionless experience. Introducing direct manipulation controls did not have immediate payoff, because they introduced new graphical representations (marks, states, tiered editing) that participants had to hold in mind alongside the narrative description of what was going on in the animation. 

However, some participants said that this was just a matter of software learning and GUI familiarization, ``\textit{More options is often always good. It’s just a matter of learning the tool. If there’s a way to have a basic view, intermediate view, expert view, that’s something I’ve liked to make it easy to find the tools you will know how to use.}" P2 said that something similar, about progressively disclosing more controls. For them, the ideal workflow would have been to start off with the prompt-only baseline, and to then move out into the precision controls of \textit{fog}. P2 said, ``\textit{The combination of the two would be to generate a first rough draft with the second version [prompt-only baseline] and then you can go through and finetune it with the first [fog].}” 

Furthermore, participants had split opinions on latency. P6 described that in an interface with many controls, where some were fully AI mediated (Tier 1 prompts), some were more direct and deterministic (Tier 3 speed sliders), and some could be combined (e.g. generating and swapping out secondary layers, passing in a path as a part of a prompt), there needed to be measures that reduced ambiguity before AI began implementing and making the user wait. P6 stated that simple things should take simple amounts of time. They wanted to know how to quickly find the threshold between what work they should do manually and what work would be easier to handoff to AI. P8 likewise wanted interface feedback on what parts of the prompt were ambiguous, complex, or would trigger more latency.

\subsubsection{Editing Patterns across Tiers}

\begin{figure}[H]
        \centering
        
        \includegraphics[width=0.8\linewidth]{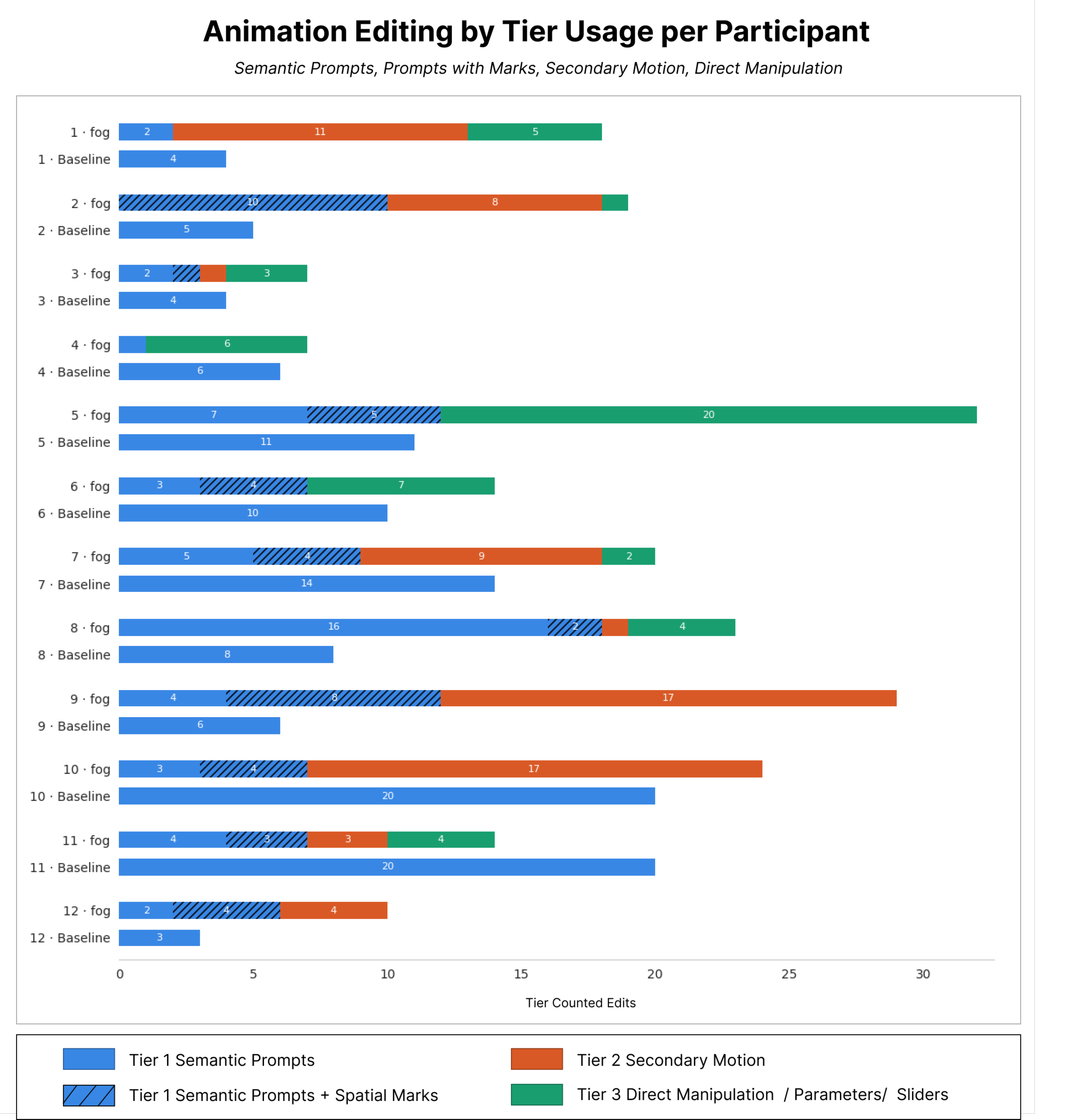} 

        \caption{Counts of editings categorized by editing tiers (T1-T3) for each participant. \textit{fog} encouraged significantly more interaction and allowed users to add spatial marks to prompts. }
        \Description{Breakdown of tier usage per participant. Each participant has two bars: one for fog and one for baseline. We can see that fog has some larger horizontal bars across its set. There is also a multivariate color bar if the user used multiple tiers of editing, most did. Some participants like P9 and P10 have expansive orange bars meaning they used the secondary motion layer.}
        \label{fig:tier-usage}
    \end{figure}

Lastly, we discussed how participants interaction patterns differed in terms of the ladder T1-T3 we described. Did participants make use of the T1) the semantic controls and prompts, T2) the swappable SecondaryMotion manner and feeling functions, and T3) the perceptual refinement controls? We depict the usage across tiers across participants in Fig.~\ref{fig:tier-usage}. Tier 1 reflected the number of times users did a semantic edit, either by passing in a prompt (with or without spatial marks) or by picking suggested motions. Tier 2 reflected the number of times users generated new \verb|SecondaryMotion| layers and tried different swappable layers. Tier 3 reflected the number of times users spent making direct manipulation or perceptual refinement actions: dropping a mark, drawing a path, editing the speed, or editing the boundaries of states. We can see from this interaction pattern chart that \textit{fog} did help participants descend the ladder from semantic to perceptual editing. In \textit{fog}, there were on average 24.33 editing interactions, and in the baseline there were on average 9.25 editing interactions. When we run a Wilcoxon signed-rank test comparing the total editing counts \textit{fog} versus baseline, we find that this difference is significant ($W $=76.5  $p$=0.0007)-- \textit{fog} does indeed drive more interaction. Part of this is by construction as \textit{fog} had more controls (Tier 2 and Tier 3). However, this shows that users did descend down the ladder of editing controls and try different granularities of control. 

Notably in Fig.~\ref{fig:tier-usage}, we can see that many prompts carried spatial marks (dashed segments in each \textit{fog} bar). Participants consistently added spatial context to their prompts to point on the canvas and draw paths. The added orange bands representing exploration of swappable SecondaryMotion layers also showed that users made use of the generation on-the-fly function to modify their motion's manner or feeling. P9 generated the most manner and feeling Secondary Motion layers, creating 17 motion functions for words like aggressively, alarmed, frightened, excitedly, bouncily, and comforted. When generated as a function, these manner and feeling functions would also expose parameters users could tweak. For example, P12 bumped up the value of \verb|Eager| after an Adverb for \verb|eagerly| was implemented. P2 appreciated that this separate endpoint for generation, because they felt it could help them work in parallel and manage the latency of the general prompting.

\subsubsection{User Study Conclusion}
Our user study hypothesized that exposing more controls would lead to a greater sense of perceived spatial control. What we found however was that baselines powered by the frontier models can do much of the heavy-lifting for animation editing, even when they are constrained and without the ability to spatially or temporally localize the prompts. They can be preferred for their immediate ease of use. However, spatial controls like pen tools and tiered editing still do provide benefits in control: they specify what words cannot and they can drive more interaction interactions in the same amount of time. The majority of participants stated that while there is some upfront cost to learning how prompts bind to GUIs, that learning could translate to more control during editing and iteration.

\section{Discussion}


We conclude by discussing what \textit{fog}'s tiered and layered approach to code generation can mean for other problems related to code generation and motion.

\subsubsection{Class abstractions as a new prompt surface for AI}

Prompts are the default for AI interaction. However, one of \textit{fog}'s ideas is to move towards abstract class definitions (\verb|PrimaryMotion|, \verb|SecondaryMotion|) rather than prompts. These code representations can define contracts for what behavior should be inherited and implemented. One of the ambitions in the paper was to write abstract classes that imposed structure but were general enough, so users could generate functions for words. The stretch goal was to give users the ability to generate an open-ended motion vocabulary. However, as we found in our technical evaluation, there was a decrease in quality once we started to combine functions that had state separate across two functions (User Study 1, R4 lost to the rest of the settings). It was more powerful to generate one monolithic function that kept the state (variables tied to the entity) altogether. While we moved to this primary monolithic function and swappable secondary layers approach in the user study, we felt that the secondary layer was diminished compared to Study 1, because some properties had moved out and into the control of the primary \verb|Verb| layer.

Figuring out what properties should go where was another challenge of system design. Should speed be owned by \verb|Verb| or \verb|Adverb|? The words ``slowly" and ``quickly" are adverbs that capture speed, but to make things monolithic, speed had to move out into \verb|Verb|. An idea from linguistics is ``verb framing" -- different languages encode manner and path in verbs differently. In some languages, the verb encodes the path and manner of the motion, in others, it does not. \textit{fog}'s hierarchical approach applies to English, but not all languages.

\subsubsection{Progressive disclosure from prompt to GUI}

One of our findings in the user study was that users want interfaces that can progressively disclose complexity. They are used to the ease of use of prompting interfaces and will take the errors if it can land them at approximately the right spot. One way to take action on this finding is to create interfaces that start with prompts but help users surface the right controls at the right time. To some degree, this has been the goal of feature-rich software for a long time. Tools like Photoshop and Fusion have contextual task bars and menus. However, these only provide just-in-time tool use. We provided a more floor-to-ceiling ladder in \textit{fog} going from language to code. While it is easy to see the appeal of progressive disclosure, it is hard to decide how many layers of abstraction should be in between. We chose to go with a primary-secondary ladder, but depending on the complexity of the scene and expertise level of the animator, that could easily be not enough.


\subsubsection{Generalizability, Future Work, Limitations}


We found during the User Study process that Heider-Simmel type animations do not always need precision controls. Hence, some of our Tier 3 controls for speed and timeline controls were underutilized. There are other domains like assembly animation and robot animation where this precision may be more necessary. One line of research that could be compelling from this work is how to design expressive robot dynamics -- for example, how do you make a Roomba move softly, or do a little dance? One could imagine that making a robotic arm move gently and characteristically will be a problem statement in the near future.

There are many other lines of future work. In our experience generating emotion functions, we noticed that there was homogeneity in the responses. Generating sadness always returned the same downward sink on a circle -- perhaps because the model was reaching for the same linguistic metaphors like drowning or bouncing for joy. As many in our related work (Herb Simon, Barbara Tversky) have pointed out, emotion and motivation are at the root of our spatial behavior. Language abstracts from our experience. LLMs have it inverted, as they use symbolic languages (English, Javascript) to produce simulation. Thus, we believe it would be interesting to probe fog for the following: homogeneity across responses and linguistic biases towards English.



Lastly, we assume there are limits to the code representations of motion and emotion in general. Code most often represents animation in terms of position, rotation, scaling, paths, and randomness. There are other representations like performance input or motion retargeting from motion capture and video that could potentially get us more realistic motion dynamics. Also, if we open up our animation engine to execute scripts with more complex code structures (if-else, while loops, helper functions) as opposed to just state and duration-based transitions, we believe we could create richer animation scenes. It was also impossible for our system to generate shape deformation or illuminance changes, which would relate to motion and emotion (e.g. flutter, shadow, sulk).



\section{Conclusion}

\textit{fog} is a layered function composition framework that composes generated primary and secondary motion layers . It introduces methods to create an open-ended vocabulary of motion spanning \verb|Verbs|, \verb|Adverbs|, and \verb|Emotions|. fog's framework in interface form allows primary motion layers to accept marks, arrows and paths -- allowing prompts to be bound to spatial context. In an technical evaluation with a laddered setup, we study how more structure in abstractions impacts expressive range during generation and composition. Lastly, to demonstrate \textit{fog}'s capabilities for storytelling and user control, we conducted a mixed-methods evaluation with novices, who animated Heider-Simmel stories to express emotion and social dynamics. We found that participants enjoyed more rapid iteration and the ability to have more tiered, granular editing in $fog$.




\bibliographystyle{ACM-Reference-Format}
\bibliography{citation}

\appendix



\end{document}